\documentclass[10pt,twocolumn,twoside]{IEEEtran}

\usepackage{graphicx}
\graphicspath{{figs/}}
\usepackage{amsfonts,amsmath,amsfonts, amssymb, amstext,bm,amsthm,color,bm}
\usepackage{algorithm}
\usepackage{shortcuts_Social}
\usepackage{booktabs}
\usepackage{multirow}
\usepackage{caption2}
\usepackage{titlesec}
\usepackage{colortbl}
\usepackage[numbers,sort&compress]{natbib}
\usepackage[bookmarks=true,colorlinks,linkcolor=black,anchorcolor=black,citecolor=black, urlcolor=black]{hyperref}  

\begin{document}

\title{Community Detection in Blockchain \\Social Networks}

\author{Sissi Xiaoxiao Wu, Zixian Wu, Shihui Chen, Gangqiang Li, and~Shengli Zhang
\thanks{This work is supported by the National Natural Science Foundation of China under Grant 61701315.}
\thanks{S. X. Wu, Z. Wu, S. Chen, G.~Li, and S. Zhang are with the College of Electronics and Information Engineering, Shenzhen University, Shenzhen, China. E-mails: 
\{xxwu.eesissi, zsl\}@szu.edu.cn, {\{1900432037, 2172262956, ligangqiang2017\}@emai.szu.edu.cn}.}
}


\maketitle

\begin{abstract}
In this work, we consider community detection in blockchain networks. We specifically take the Bitcoin network and Ethereum network as two examples, where community detection serves in different ways. For the Bitcoin network, we modify the traditional community detection method and apply it to the transaction social network to cluster users with similar characteristics. For the Ethereum network, on the other hand, we define a bipartite social graph based on the smart contract transactions. A novel community detection algorithm which is designed for low-rank signals on graph can help find users' communities based on user-token subscription. Based on these results, two strategies are devised to deliver on-chain advertisements to those users in the same community. We implement the proposed algorithms on real data. By adopting the modified clustering algorithm, the community results in the Bitcoin network is basically consistent with the ground-truth of betting site community which has been announced to the public. At the meanwhile, we run the proposed strategy on real Ethereum data, visualize the results and implement an advertisement delivery on the Ropsten test net.
%
%
\end{abstract}


%
\IEEEpeerreviewmaketitle

{\section{\large Introduction}\label{sec:intro}}
\IEEEPARstart{E}{ver} since Satoshi Nakamoto's Bitcoin white paper in the year of $2008$\cite{nakamoto2008bitcoin}, blockchain has been launched in many areas such as banking\cite{guo2016blockchain}, network security, supply-chain management\cite{saberi2019blockchain}, internet-of-things (IoT)\cite{conoscenti2016blockchain}, financial cryptocurrency\cite{miraz2018applications}, serving as a decentralized ledger. Recently, governments in different countries begin to pay a huge attention to blockchain and put technologies involving blockchain into the cutting edge. In this work, instead of treating blockchain as a ledger, we try to study blockchain from a social media perspective. Specifically, we define the blockchain network as a decentralized social network, based on which we try different algorithms to analyze users' relationship underlying the ledger records. Our final goals are two-folded. First, we define different types of social networks for both Bitcoin and Ethereum. Second, by discovering users' clusters in the defined social graph we analyze users' behavior in both networks, and especially try to deliver advertisements in the Ethereum network.

Social network data is valuable as we can mine users' preferences from it and thus explore potential marketing. In traditional centralized social network, data is stored in a fusion center which is owned by the platform. Therefore, the platform monopolizes all data mining applications to make a big fortune. The blockchain network, however, decentralizes data among users and allows each user in the blockchain network fully access the whole piece of data and develops its own applications. Moreover, traceability ensured by the blockchain endorses the quality of the data, which further improves the efficiency of the data mining applications. The above nice properties make the blockchain as a social network promising in the future.

In this work, according to the way of recording the ledgers, Bitcoin and Ethereum are respectively defined as different kinds of social networks. In Bitcoin, every user can generate multiple private-public key pairs and the only purpose of transaction is to send BTC coins. The public key (also known as the address) is visible in the transaction block, as either the transaction input or the transaction output. Multiple input addresses and multiple output addresses may exist in the same Bitcoin transaction. In this context, one important pre-processing task of analyzing the ledger data in Bitcoin is to associate those addressees which belong to the same user and group them into a super-node in the social graph. This kind of operation is usually referred to ``common spend" or ``change address" heuristics.
Based on the pre-processing results, we define the Bitcoin social network as an undirected graph where each super-node corresponds to a node in the graph and the edge weight of any two nodes in the graph is defined by historical coin-based transactions between any two super-nodes. Then we propose a specific clustering algorithm, which originated from the spectral clustering algorithm, to the Bitcoin social graph to find communities.

Ethereum, published in $2014$ by Vitalik Buterin and launched in $2015$, is the world's leading programmable blockchain
as it has added a self-enforcing piece, i.e., the so called \textit{smart contract}, which ensures a coordinated and enforced agreement between network participants by means of an organization or to create tokens\cite{buterin2014next}. In Ethereum, a transaction has only one input address and one output address and thus we can bypass the super-address pre-processing. Unlike Bitcoin, users in Ethereum interact with each other by not only a direct ETH coin transaction but also the smart contract transaction. In this work, we focus on the smart contract transactions and define the Ethereum social network as a bipartite social graph. Particularly, we are interested in those smart contract transactions specific for initial coin offering (ICO) events. Based on their bipartite graph, we introduce an effective community detection algorithm for low-rank signals to group users into different clusters\cite{wai2018community}. These results can be further used for other purposes. For example, in both Bitcoin and Ethereum networks, if one person creates two user accounts, it is generally difficult to associate those accounts to the same person due to pseudo-anonymity. Our community detection algorithms may help do the association by analyzing the accounts' preferences, given that two accounts for the same user should share similarities. Also, the communities results may also be used to label users' potential preferences and thus provide a targeted referral service in blockchain.

 %

\subsection{Related work}
There has been a branch of works in the literature on analyzing the Bitcoin transaction data and most of them focus on two issues: anonymization and de-anonymization. The review on these two issues can be found in Refs. \cite{shentu2015research, conti2018survey, reid2013analysis, ron2013quantitative}.
Therein, shared coin and send mixers in Refs. \cite{barber2012bitter, maxwell2013coinjoin,maxwell2013coinswap,bonneau2014mixcoin,yanovich2016shared} are two basic anonymization approaches, while other approaches such as fair exchange\cite{barber2012bitter}, transaction remote release\cite{shentu2015transaction} and zero cash\cite{sasson2014zerocash} are also popular. In this work, our main focus is on de-anonymization. In the literature, there are many approaches for implementing de-anonymization. For example, an early work in Ref. \cite{barber2012bitter} first brought up the notion of ``change address" and people realized that one can use the heuristics to associate addresses that involving common spending and one-time change. This approach is widely used as the first step to process Bitcoin data\cite{ermilov2017automatic, fleder2015bitcoin,harrigan2016unreasonable,meiklejohn2013fistful,remy2017tracking,chang2018improving,androulaki2013evaluating}.
There are also more advanced approaches. In Refs. \cite{kaminsky2011black,biryukov2015bitcoin,zhu2017mining}, the authors tried to de-anonymize user's identity of Bitcoin by linking the Bitcoin address with IP address. Ref. \cite{spagnuolo2014bitiodine} summarized prior approaches of clustering addresses and implemented a modular framework to group users and classify them with different labels.
Sometimes, off-chain information is also useful. For example,  in Refs. \cite{reid2013analysis,ermilov2017automatic} the authors proposed to use off-chain information to guide different clustering models with a purpose of reducing the algorithm complexity. Ref. \cite{dupont2015toward} and Ref. \cite{monaco2015identifying} proposed novel methodologies for analyzing Bitcoin users based on the observation of Bitcoin transactions over time.

In this work, we try to deanonymize a blockchain network by finding users' communities. Some prior works have been done for Bitcoin. The idea of treating Bitcoin as a social network has appeared in Ref. \cite {somin2018social}. Therein, people usually used the notions of ``user graph" and ``transaction graph", and some analysis based on these two graphs has been elaborated in Ref. \cite{reid2013analysis}. Ref. \cite{Moser} studied the anonymity for Bitcoin by analyzing the transaction graph with the help of public data. 
Ref. \cite{goldsmith2019analyzing} studied how different features of the date influence communities results on the ``transaction graph" based on the ground truth of some known hack subnetworks. Authors in Ref. \cite{pham2016anomaly} and Ref. \cite{hossain2018community} extracted various features from these two graphs and pointed it out that features on the graph are crucial for the analysis results. Ref. \cite{remy2017tracking} showed that a two-party community detection based on normalization mutual information could be used to re-identify users in Bitcoin.
Our community detection approach on the Bitcoin data is based on the above works. Specifically, our approach has the following properties: 1) we study the ``user graph" based on the super-address which is associated with addresses that involve common spending and one-time change; 2) we use historical BTC coin amount as the key feature to perform the community detection algorithm; 3) a modified clustering method is proposed for the Bitcoin social graph to find communities. 

The above approaches are effective for Bitcoin while it is difficult to be directly carried forward to the Ethereum network, as Ethereum is mechanically different from Bitcoin. In Ethereum, there are two types of accounts: externally-owned accounts (EoAs) and contract accounts (CAs). 
In Refs. \cite{klusman2018deanonymisation, chan2017ethereum}, the authors pointed it out that existing methods such as discovering IP addresses and Bitcoin addresses clustering usually do not fit the Ethereum network due to the differences between both networks in the volatility of entry nodes and the way transactions are handled.
Some works tried to use traditional clustering methods for Ethereum, such as support vector machine (SVM) and k-means in Ref. \cite{brinckman2019techniques}, the standard k-means in Refs. \cite{petrov2019identification, huang2018soc}, long short-term memory (LSTM) and cnonvolutional neural network (CNN) models in Ref. \cite{huang2018soc}, affinity propagation k-medoids in Ref. \cite{norvill2017automated},  k-means clustering, agglomerative clustering, Birch clustering in Refs. \cite{payette2017characterizing, sun2019ethereum} and Neo4j in Ref. \cite{chan2017ethereum}. However, they basically equally treat EoAs and CAs as nodes in the transaction graph\cite{petrov2019identification, chan2017ethereum, sun2019ethereum}, while these two types of accounts are essentially different. Some works in Refs. \cite{norvill2017automated, linoyexploring} also used side information to analyze the on-chain transactions. Therein Ref. \cite{linoyexploring} utilized the smart contract codes to analyze the smart contract nodes in the transaction graph.
It is worth mentioning that our community detection approach in a bipartite graph differs from traditional one in a bipartite graph as we treat nodes on both sides of the bipartite graph as two parties of nodes, therefore utilizing the connections between those two parties to cluster nodes in one party, while the traditional approach usually treats all the nodes in both sides equally and find communities over all the nodes in both sides \cite{barber2007modularity,alzahrani2016community,zhou2018novel}.
Our work on Ethereum data differs from the above work in five aspects: 1) we analyze the on-chain data without any off-chain side information; 2) we separately treat EoAs and CAs as different nodes and put them on two sides of a bipartite social graph; 3) we target on smart contract transactions which involve ICO events; 4) we apply a novel low-rank community detection algorithm based on graph signal processing (GSP) on the bipartite graph; 5) we utilize the clustering results to deliver on-chain advertisement in Ethereum. 

The rest of the paper is organized as follows. In Section \ref{sec:bitcoin}, we show an example to define the Bitcoin social graph and then apply a novel clustering method on this graph. Based on this result, Section \ref{sec:Ethereum} demonstrates the difference between the Bitcoin social graph and the Ethereum social graph. We then define the bipartite Ethereum social graph and utilize a particular method to find communities on it. Simulation results are illustrated in Section \ref{sec:numeric} where both Bitcoin data and Ethereum data are analyzed and compared. Moreover, we also test the on-chain advertisement for Ethereum based on the clustering results. We conclude and further discuss with simulation results in Section \ref{sec:conclusions}.

\begin{figure}[t!]
\centering{ \includegraphics[width=8cm] {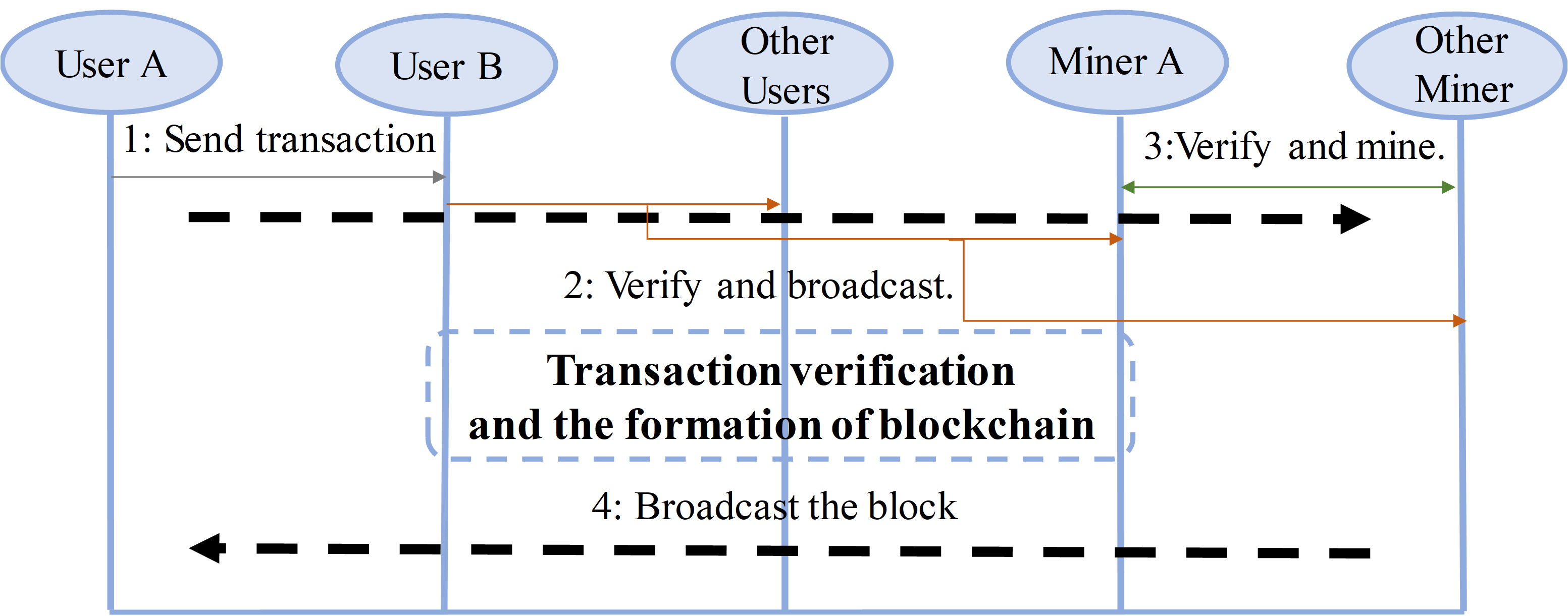}}
\caption{\small The consensus process in Bitcoin network.}
\label{fig:consensus}
\end{figure}

{\section{Bitcoin Transactions Analysis} \label{sec:bitcoin}}
To start our work, we first introduce how Bitcoin verifies and records transactions.
As shown in Fig.~\ref{fig:consensus}, in a Bitcoin network, when one node initiates a transaction, the transaction information will be signed and packaged, thus broadcasting to other nodes. Those nodes who receive the transaction will verify its legality and then help broadcast the verified transaction message.
During the process of broadcasting, some of the received nodes are miners, who not only bear the responsibility of broadcasting transactions, but also undertake the task of ``mining". The miner who has successfully mined by solving a difficult mathematical problem will get the right to write the ledger and add all transactions they have verified. When most of the nodes in the entire network agree on the same transactions, these transactions are recorded in the block.

\subsection{The Bitcoin Social Network}
\begin{figure}[t!]
\centering{ \includegraphics[width=8cm] {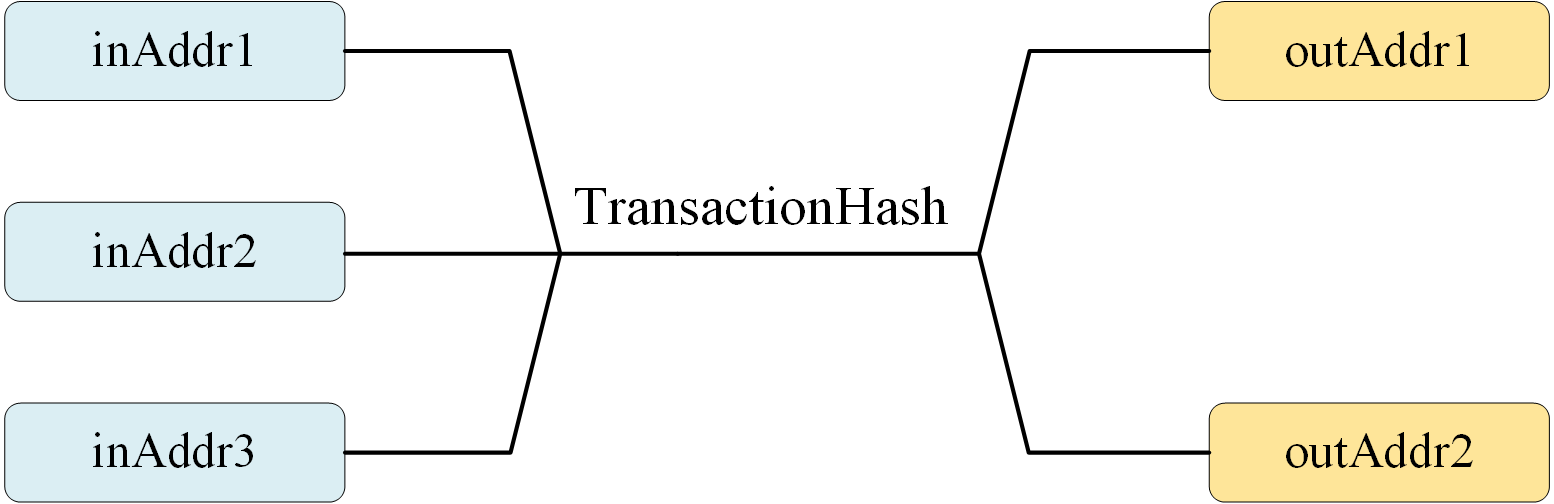}}
\caption{\small The Bitcoin UTXO model: ``inAddr" and ``outAddr" are the abbreviations of the input and output addresses. We use the ``transactionHash" to denote the transaction including these addresses. }
\label{fig:utxo}
\end{figure}

As mentioned in the previous section, the only purpose of transaction in Bitcoin is to transfer BTC coins. To achieve this goal, each user generates a key pair (represented as the addresses in the transaction) to join the Bitcoin network and transfer BTC coins based on the so-called unspent transaction output (UTXO) model\cite{nakamoto2008bitcoin}. UTXO can be seen as an abstraction of electronic money, representing a chain of ownership implemented as a chain of digital signatures.
In Fig. \ref{fig:utxo}, we show a basic structure of a Bitcoin transaction where we can find multiple addresses in the input and output fields. Every address could contain multiple UTXO, wherein UTXO in the input addresses are consumed while in the output addresses are created.

A basic idea to define the Bitcoin social network is to let each address one-to-one correspond to a node in the social graph. However, a crucial problem here is that one user usually possess multiple addresses. Given that there are so many addresses in the blockchain, the dimensionality of the social network could be huge. To reduce the size of the graph, we try the following way to associate multiple addresses (key pairs) to a super-address in the social graph: 1) multiple input addresses of a transaction; 2) those bitcoin users that have a common change address. More details for such operations will be shown in the numerical experiments. After such a pre-processing, we define a social network where each node corresponds to a processed super-address.

\begin{algorithm}[t!]
	\textbf{Input}: a set of nodes {$\bm{V} = \{ {v}_1,\cdots,{v}_n \}$}, the number of clusters $k$ and weight matrix $\bm{W}$ for all nodes in $\bm{S}$ \\
	\textbf{Step 1:} Define $\bm{D}$ to be the diagonal matrix whose $(i, i)$-element is the sum of $\bm{W}$'s $i$-th row. Letting $\bm{L} = \bm{D} -\bm{W}$, construct a matrix $\bm{\bar L} =\bm{D}^{-1/2}\bm{L}\bm{D}^{-1/2}$.\\
	\textbf{Step 2:} Find $\bm{x}_1,\bm{x}_2,...,\bm{x}_k$, the $k$ largest eigenvectors of $\bm{\bar L}$ (chosen to be orthogonal to each other in the case of repeated eigenvalues), and form the matrix $\bm{X} = [\bm{x}_1,\bm{x}_2,...,\bm{x}_k]$ by stacking the eigenvectors in columns.\\
	\textbf{Step 3:} Form the matrix $\bm{Y}$ from $\bm{X}$ by renormalizing each of $\bm{X}$'s rows to have unit length$(i.e., \bm{Y}_{ij} = \bm{X}_{ij}/(\sum_{j}{\bm{X}^{2}_{ij}})^{1/2})$.\\
	\textbf{Step 4:} Treating each row of $\bm{Y}$ as a point in $\mathbb{R}^{k}$ , cluster them into $k$ clusters via $k$-means or any other algorithm that attempts to minimize distortion.\\
	\textbf{Step 5:} Finally, assign the original point $\bm{v}_i$ to cluster $j$ if and only if row $i$ of the matrix $\bm{Y}$ was assigned to cluster $j$.\\
	\textbf{Output}: Partition nodes in $\bm{V}$ into $k$ communities.
	\caption{Clustering the Bitcoin social graph}\label{alg:1}
\end{algorithm}
\subsection{Community Detection for Bitcoin}
To well define the graph, we need to specify the edge weight between any two nodes (super-addresses). In fact, the edge weight between any two nodes could be defined following criterion in Refs. \cite{goldsmith2019analyzing, pham2016anomaly, hossain2018community}. Herein, we extract features from the total transaction amount and set them as the edge weight. Then, we run a clustering method which is modified from the spectral clustering algorithm\cite{ng2002spectral, wang2019improvement} to cluster this Bitcoin social graph. Specifically, we denote the social graph by $\bm{G(V,E)}$,
where $\bm{V}$ denotes the user nodes $(v_1, v_2, ... v_n)$ and $\bm{E}$ the edges. We let $\bm{W}$ denote the weight matrix where each entry
$w_{ij}$ is the edge weight between node $v_i$ and node $v_j$.
In this paper, we let $a_{ij}$ be the historical total transaction amount between node $i$ and node $j$ and
\[
w_{ij} = a_{ij}/\max_{i, j} \{a_{ij}\}.
\]
Apparently, we will thus have an undirected graph with $w_{ij}=w_{ji}$. 
We then apply the following clustering algorithm in Algorithm \ref{alg:1} to find communities in the graph.

{\it Remark 1:}
It is worth noting that the above clustering algorithm is quite similar to the well known spectral clustering algorithm except for that the similarity matrix is replaced by the weight matrix in our algorithm. This modification is meaningful since we re-define the ``similarity" as that two nodes have a significant transaction relationship. This redefinition can help cluster users that have more connections with each other.

{\section{The Ethereum Social Network} \label{sec:Ethereum}}
As a blockchain network, Ethereum is different from Bitcoin in many aspects\cite{buterin2014next, wood2014ethereum,antonopoulos2018mastering}. Particularly, Ethereum is not only a platform for providing ETH coin transactions, but also a programming language that enables users to build and publish distributed applications via the smart contract.
In Ethereum, each user generates a pair of asymmetrically encrypted public key and private key to join the network. Each public key could be considered as a node in our Ethereum social network.
In Ethereum, there are two types of accounts: externally-owned accounts (EoAs) and contract accounts (CAs).
EoAs are considered as individual users in the external world while CAs are the contracts that could connect EoA users. Both EoAs and CAs are presented by unique hash addresses.

\begin{figure}[t!]
	\includegraphics[width=8cm]{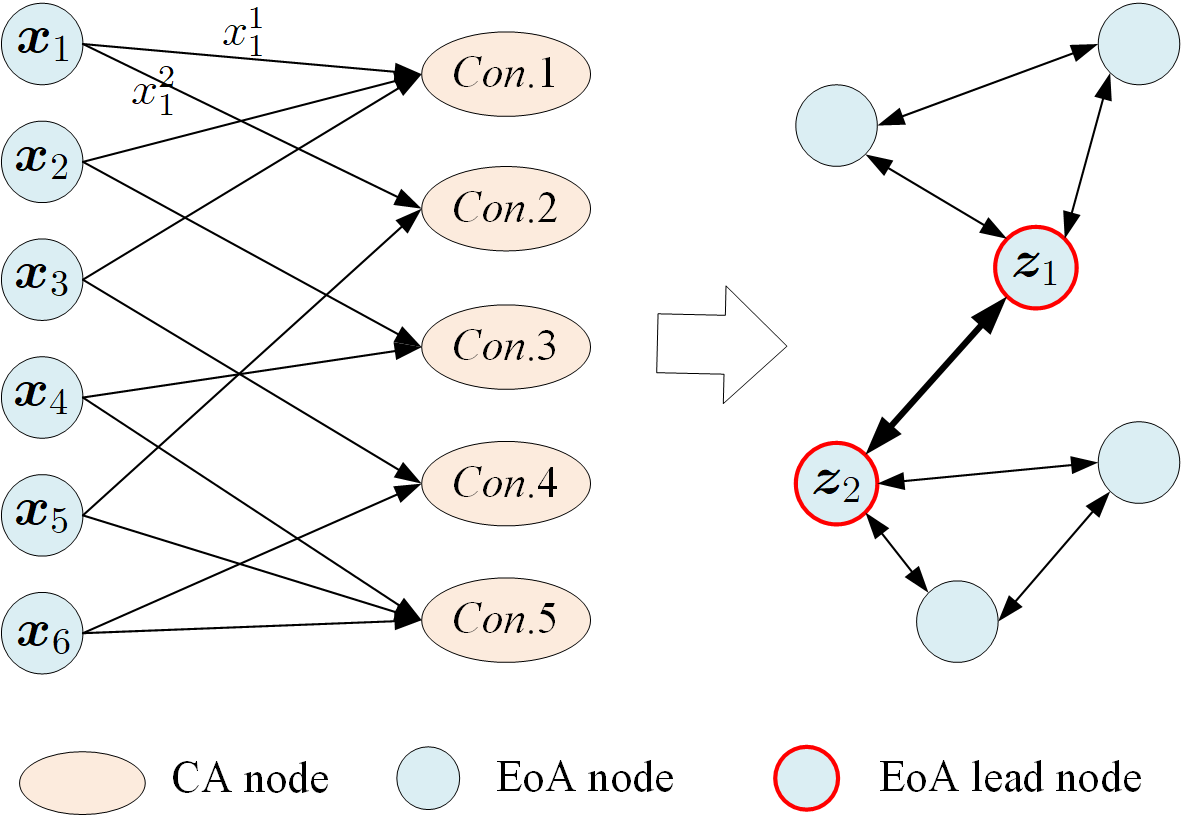}
	\centering
	\caption{{\small Construct a social network based on different ICO tokens: ${x}_i^t$ represents user $i$'s balance for token $t$.}}
	\label{a_bipartite_network}
\end{figure}

According to the properties of Ethereum, we define its social network as a bipartite graph, where EoA nodes and CA nodes are put into two sides of
the graph; seen in Fig.~\ref{a_bipartite_network}. 
Each EoA node has their attention on different CA nodes. 
{For example, in Fig.~\ref{a_bipartite_network} (left), supposing that we have $N$ EoA nodes and $T$ CA nodes in the graph, for an EoA node $i \in \{1,...,N\}$, we define ${\bm x}_i = [x_i^1, x_i^2,...,x_i^T]$ and $x_i^t$ is EoA node $i$'s attention on CA node $t$. }
Herein, the CA nodes could be any smart contract in Ethereum.
{A typical example could be a token created by an ICO event, where a possible choice of ${x}_i^t$ could be user $i$'s transaction amount on token $t$. }
 We remark that in this paper we use ICO events and token transaction amounts as features to define the bipartite graph, while it actually could be any other type of smart contracts which connect EoA nodes.

\begin{algorithm}[t!]
	\textbf{1: Input}: Graph signals $ {\bm y}_{t=1}^T $ ; desired number of clusters $K$. \\
	\textbf{2:} Use ${\bm y}_{t=1}^T$ to compute the sample covariance $ \hat{\bm{C}}_{x} $ as in (5).\\
	\textbf{3:} Find the $K$ eigenvectors to $ \hat{\bm{C}}_{x} $ associated with the largest $K$ eigenvalues. Denote the set of eigenvectors as $ \hat{\bm{P}}_{K} \in   \mathbb{R}^{{N} \times K}$.\\
	\textbf{4:} Perform $K$-means clustering\cite{von2007tutorial}, which optimizes:
	\begin{equation} 
	\min \limits _{    {\cal C}_1,...,{\cal C}_K}   {\sum_{i=1}^{K}}   {\sum_{j \in {\cal C}_i}}    { \| \hat{\bm{p}}_{j} - \frac{1}{|{\cal{C}}_{i}|}\sum_{q \in {\cal{C}}_{i}} {\hat{\bm {p}}_{q}\|_{2}^{2}}} \quad  s.t.  \quad  {\cal{C}}_{i} \in V \notag
	\end{equation}
	\textbf{\quad} where $\hat{\bm{p}}_{j} := [\hat{\bm{P}}_{K}]_i,: \in \mathbb{R}^{ K} $.Let the solution be $ \hat{\cal C}_1,...,\hat{\cal C}_K $.\\
	\textbf{5:Output}:   Partition of ${\bm V}$ into $K$ communities,$ \hat{\cal C}_1,...,\hat{\cal C}_K $.
	\caption{Community detection from low-rank excitation}\label{alg:2}
\end{algorithm}

\subsection{Algorithms and Strategies}\label{sec:alg}
Now our purpose is to perform a community detection on this bipartite graph and group all EoA nodes into clusters.
In this subsection, we adopt the low-rank community detection algorithm in Ref. \cite{wai2018community} to cluster the EoA nodes in the bipartite graph. The idea is to assume that all EoA nodes form a low-rank social sub-graph, where some lead EoA nodes will decide other nodes' attention on CA nodes. In this spirit, we partition the EoA node set of the bipartite graph into subsets with high edge densities. This could be done by applying a clustering algorithm on the low-rank output covariance matrix of the observed graph signal at EoA nodes. To proceed it, we regard this community detection problem as a problem of GSP, wherein the input of the graph ${\bm z} \in \mathbb{R}^{R} $ is on the EoA lead node (see Fig.~\ref{a_bipartite_network}) and it goes through a filter ${\cal H}({\bm S})$ :
\begin{equation}
\mathcal{H}(\bm{S}): ={\sum}_{\ell=0}^{L-1}h_{\ell}\bm{S}^{\ell}=\bm{V}({\sum}_{\ell=0}^{L-1}h_{\ell}{\bm \Lambda}^{\ell})\bm{V}^{H}
\end{equation}
where $\bm{S}$ is the graph Laplacian matrix, $L$ is the degree of the filter and $R$ is the number of lead node.
The output signal ${\bm x} \in \mathbb{R}^{N} $ is defined on all the EoA nodes and it is generated by
\beq \label{eq:x}
{\bm x} = \mathcal{H}(\bm{S}){\bm z}\eqs.
\eeq
Herein, ${\bm V}$ and ${\bm \Lambda}$ are from a SVD decomposition of ${\bm S}$. The above equation means that in our graph model, the opinion of the EoA lead node decides all nodes' status. Based on the above model, the graph signal observed at all the EoA nodes can be expressed as
\beq\label{yt}
\bm{y}^{t}=\bm{x}^{t}+\bm{w}^{t}\ \mathrm{and}\ \bm{x}^{t}=\mathcal{H}(\bm{S})\bm{z}^{t}, \quad t=1,...T
\eeq
where $\bm{y}^{t}$ is the observation of the graph signal which represents EoA nodes' attention on CA node $t$ in our problem setting and ${\bm w}^t  \sim {\cal N}(0, {\sigma}^2  {\bm I}) $ is the noise. 
Notice that the input signal ${\bm z}^t$ is applied on only a subset
of $R$ EoA lead nodes and thus the number of variations in the excitation signal is limited to $R$ mode.  To further explore the graph structure, we let
\beq
\bm{C}_{z}=\mathbb{E}[\bm{z}^{t}(\bm{z}^{t})^{\mathsf{T}}]=\bm{BB}^{\mathsf{T}},
\eeq
where ${\bm B} \in \mathbb{R}^{N \times R} $ with $R < N$.
Then, we can recover the community structure in ${\bm S}$ by applying Algorithm \ref{alg:2} on the empirical sampled covariance of the observed signal $\bm{y}^{t}$:
\beq \label{eq:Cy}
\hat{\bm{C}}_{x}=(1/T){\sum}_{t=1}^{T}\bm{y}^{t}(\bm{y}^{t})^{\mathsf{T}},
\eeq
which is an estimate of $\bm{C}_{x}=\mathcal{H}(\bm{S})\bm{BB}^{\mathsf{T}}\mathcal{H}(\bm{S})^{\mathsf{T}}$.

\begin{figure}[t]
	\centering
	\includegraphics[width=8cm]{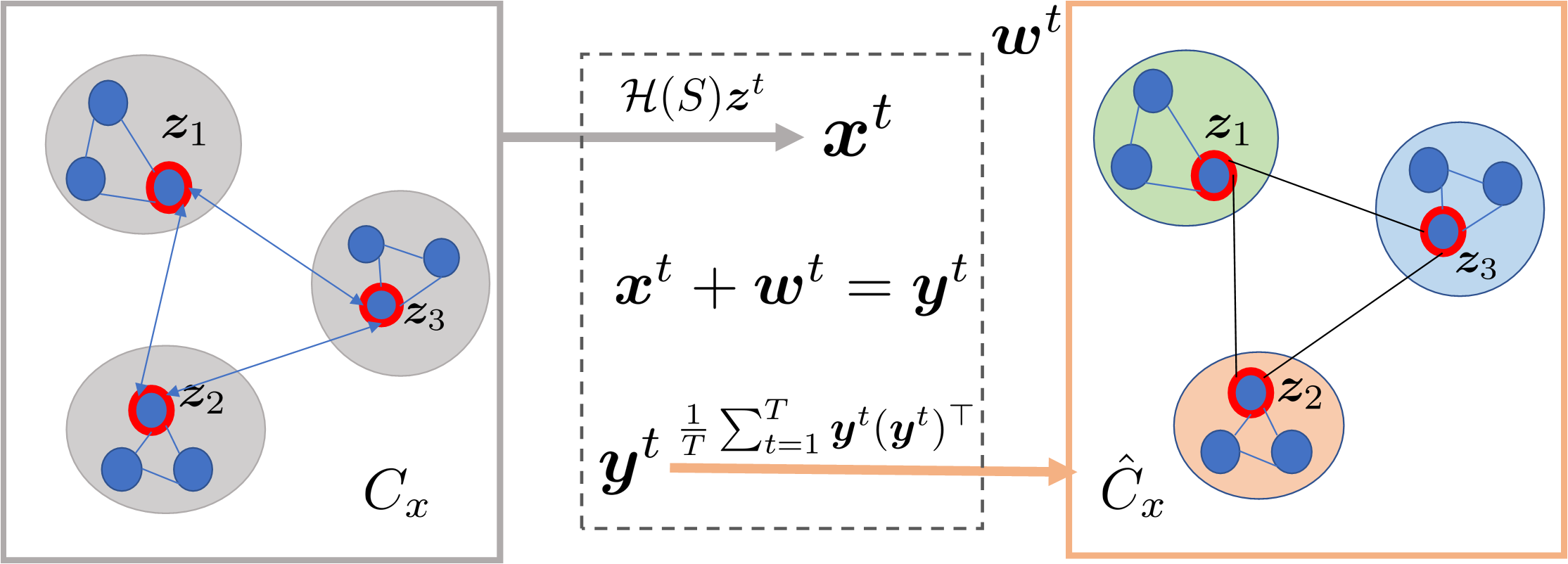}
	\caption{\small An algorithm model for Ethereum.}
	\label{fig:algorithm_model}
\end{figure}

An illustration of the algorithm model for Ethereum is depicted in Fig. \ref{fig:algorithm_model}. 
In practice, $\hat{\bm{C}}_x$ could be obtained by observing the graph signals from many instances $t$. For example, in the Ethereum social network, we consider each instance $t$ as one CA node in the bipartite graph. Thus, we obtain the graph signal ${\bm y}^{t}$ by observing EoA nodes' attention on CA nodes and utilize it to detect communities of EoA nodes. For example, we could consider users' transaction amount on different tokens as their attention on such tokens. 
\begin{figure}[t!]
	\includegraphics[width=8cm]{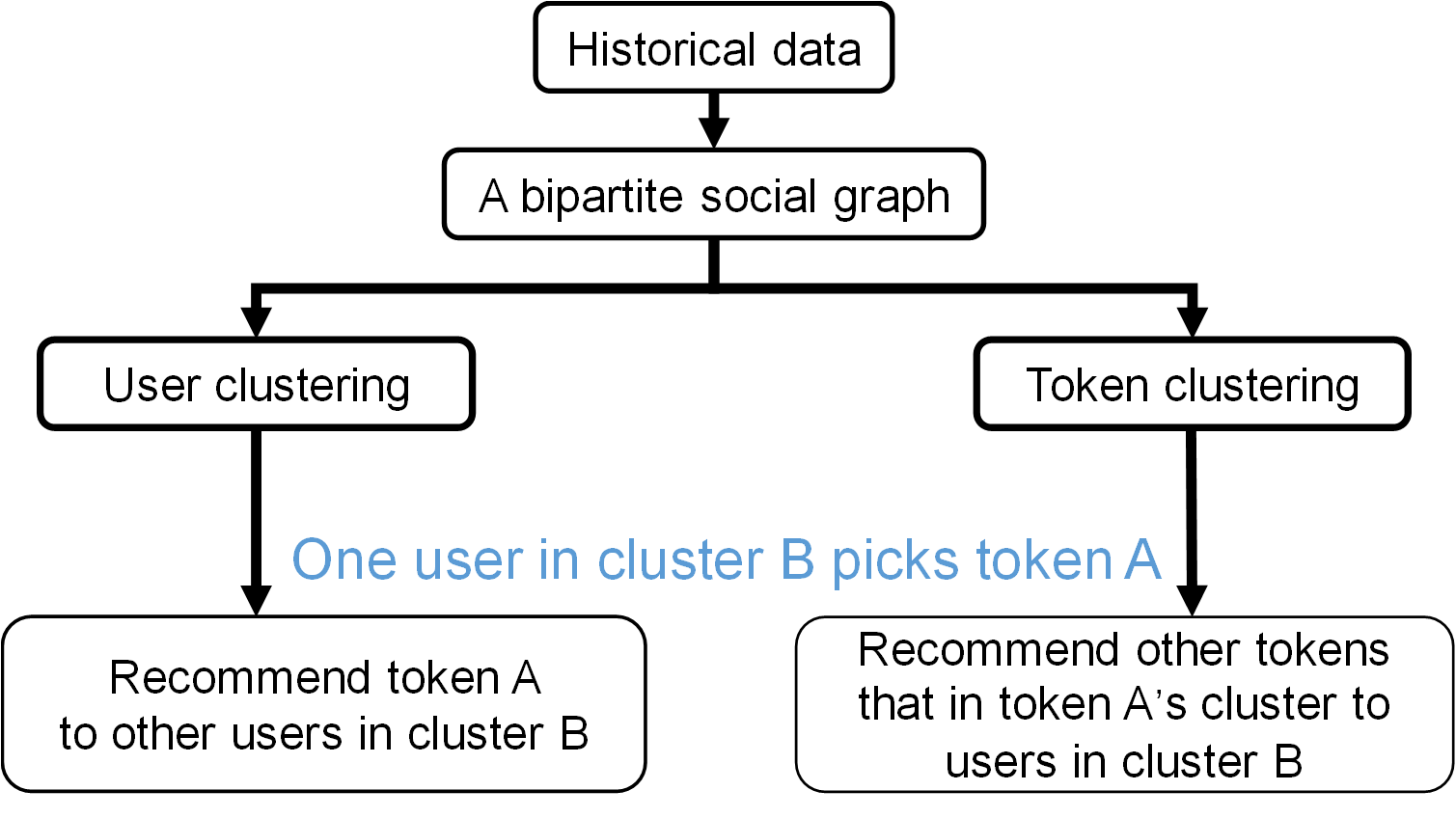}
	\centering
	\caption{\small A flow chart for the recommendation system.}
	\label{F:recom}
\end{figure}
\begin{figure}[t!]
	\centering
	\includegraphics[width=8cm]{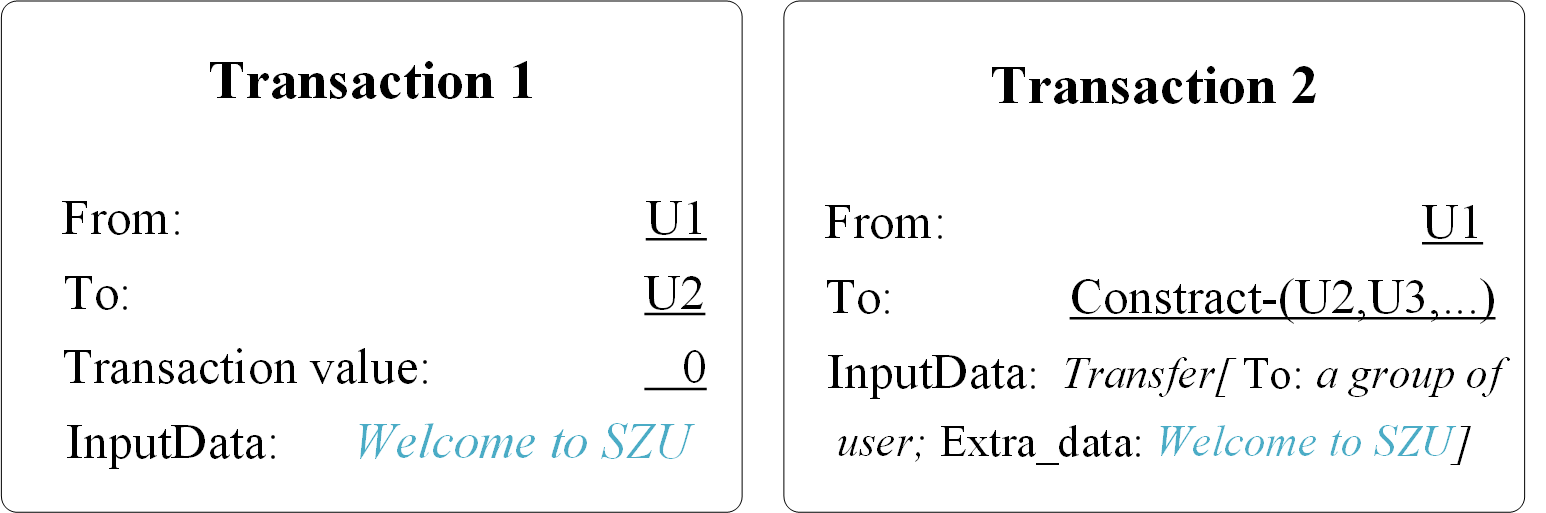}
	\caption{\small Coin transaction (left) and smart contract transaction (right). Ui denotes User $i$.}
	\label{fig:Eth_transaction}
\end{figure}
\subsection{Group Tokens by Users Subscription}
In previous discussion, we discuss how to find communities of EoA nodes by using graph signal processing. In fact, this process can be reversed to cluster tokens by users' subscription. That is, we exchange the position of the EoA nodes and CA nodes and observe the graph signal at each CA node, which represents all EoA nodes' attention
on a specific CA node. This equals to transposing $\hat{\bm{C}}_{x}$ in Algorithm \ref{alg:2} and performing the same clustering method on the new $\hat{\bm{C}}_{x}$. Numerical results are shown in Section \ref{Ethereum_Data}.

At the end of this part, two remarks are in order. First, the relationship between user clustering and token clustering is analog to that between the user-based collaborative filtering and the item-based collaborative filtering. Second, the token clustering result can also be used to recommend tokens to EoA nodes. A flow chart for the recommendation system is shown in Fig. \ref{F:recom}. We will introduce the detailed recommendation process in the next subsection.

\subsection{Advertisement Strategies}
The proposed community detection algorithms can help find EoA users sharing the same interests on CA nodes (by user clustering), as well as CA nodes favored by groups of EoA nodes (by token clustering). We then discuss 1) how to deliver advertisements to a user whose community members have shown interest on a specific token, and 2) how to recommend other tokens to a user who has shown interest on a specific token. Specifically, we may resort to the ``InputData" field in the transaction script to serve our purpose. In Fig.~\ref{fig:Eth_transaction}, we show two types of transactions in Ethereum. Transaction $1$ is an ETH coin transaction (left) and Transaction $2$ is a smart contract transaction (right). For both transactions, there is an ``InputData" field in the script which can be used to run functions or send messages.
We therefore design two on-chain advertisement strategies. One approach is to send a small amount ETH coin (could be zero) to the target user and attach a recommendation message in the ``InputData"  field in this coin transaction. This implementation can only be done in a one-to-one manner and one has to cost some gas to send the message. Another approach is similar to the so-called ``airdrop"\cite{BountyOne,di2019collateral}, wherein new ICO project distributes part of their tokens for free to a community to advertise their ICO project. ``Airdrop" could be done via smart contract in a group message manner and no extra ETH coin is consumed except for the gas. Notice that to successfully deliver the message, one has to negotiate with the wallet company to register their token in the target user's list. Otherwise, users can not see the new-added token, as well as the advertisement message. We remark that our design differs from the original
``airdrop" in two aspects: 1) we have resorted to a community detection to target potential users; 2) we utilize the ``InputData" field to send the advertisement message.
\begin{figure}[t!]
	\centering{ \includegraphics[width=8cm] {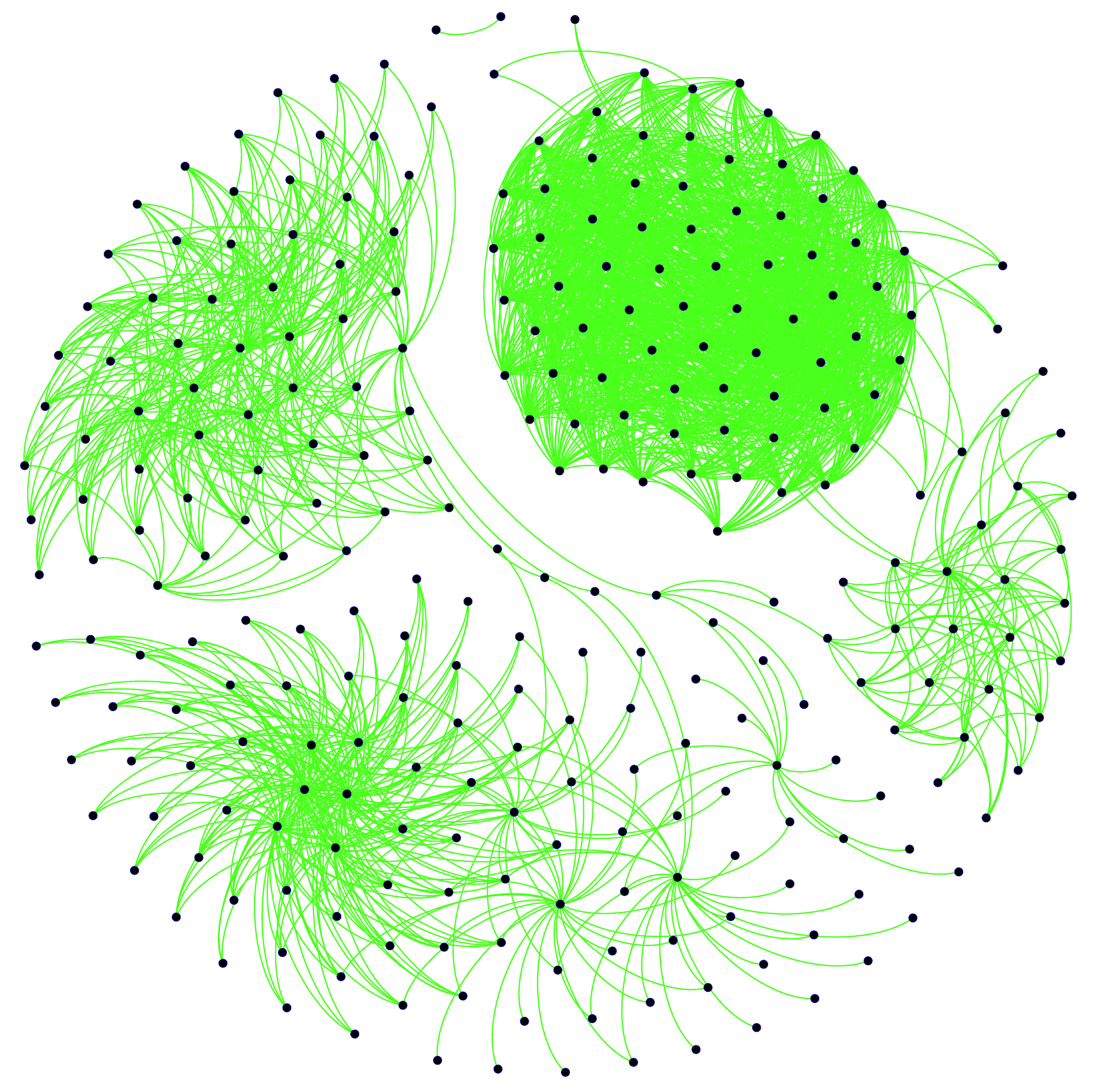}}
	\centering\caption{\small Graph representation after associating the common spend addresses.}
	\label{fig:multiaddr}
\end{figure}
\begin{figure*}[t!]
	\centering
	\includegraphics[width=8.5cm] {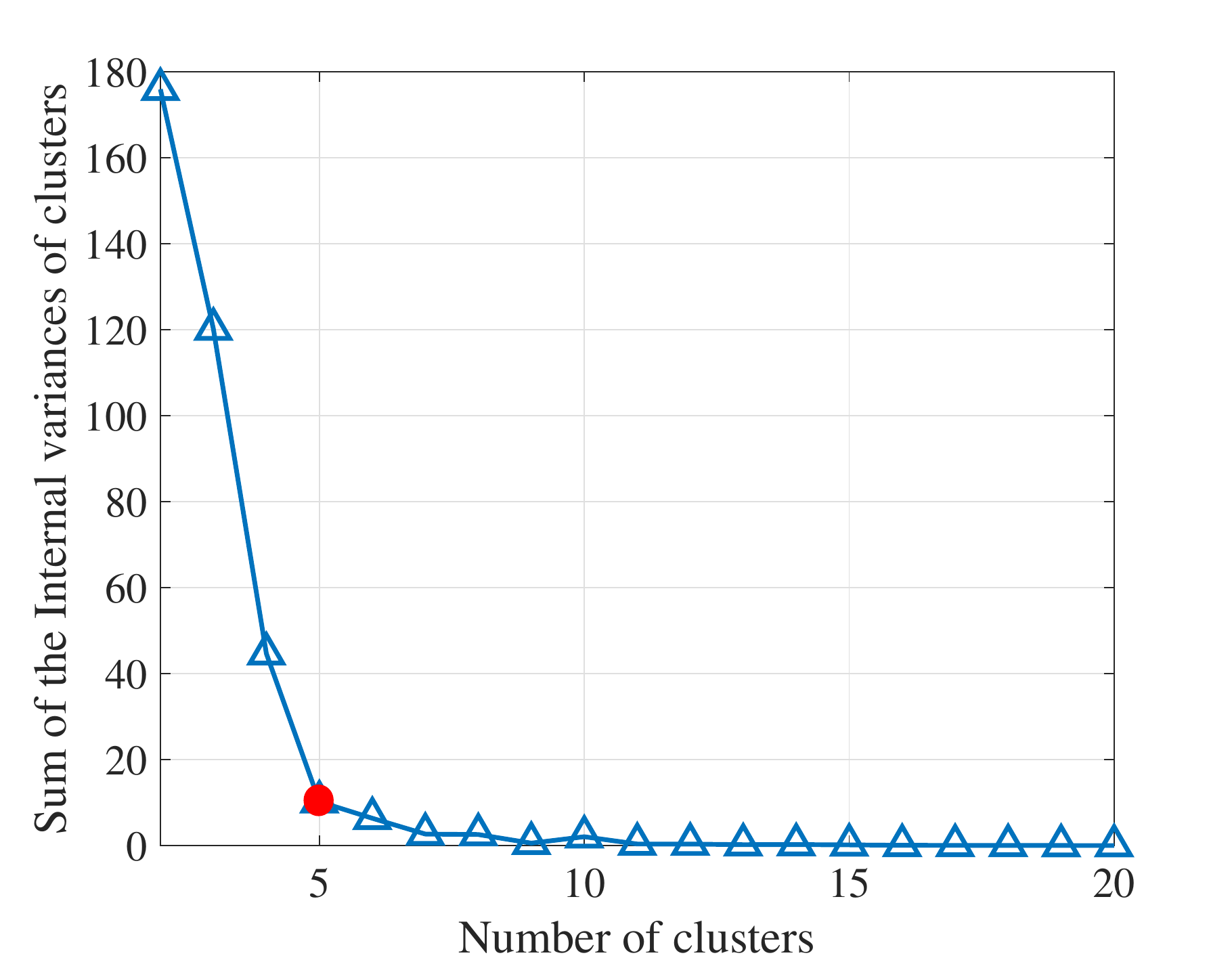}
	\includegraphics[width=8.5cm] {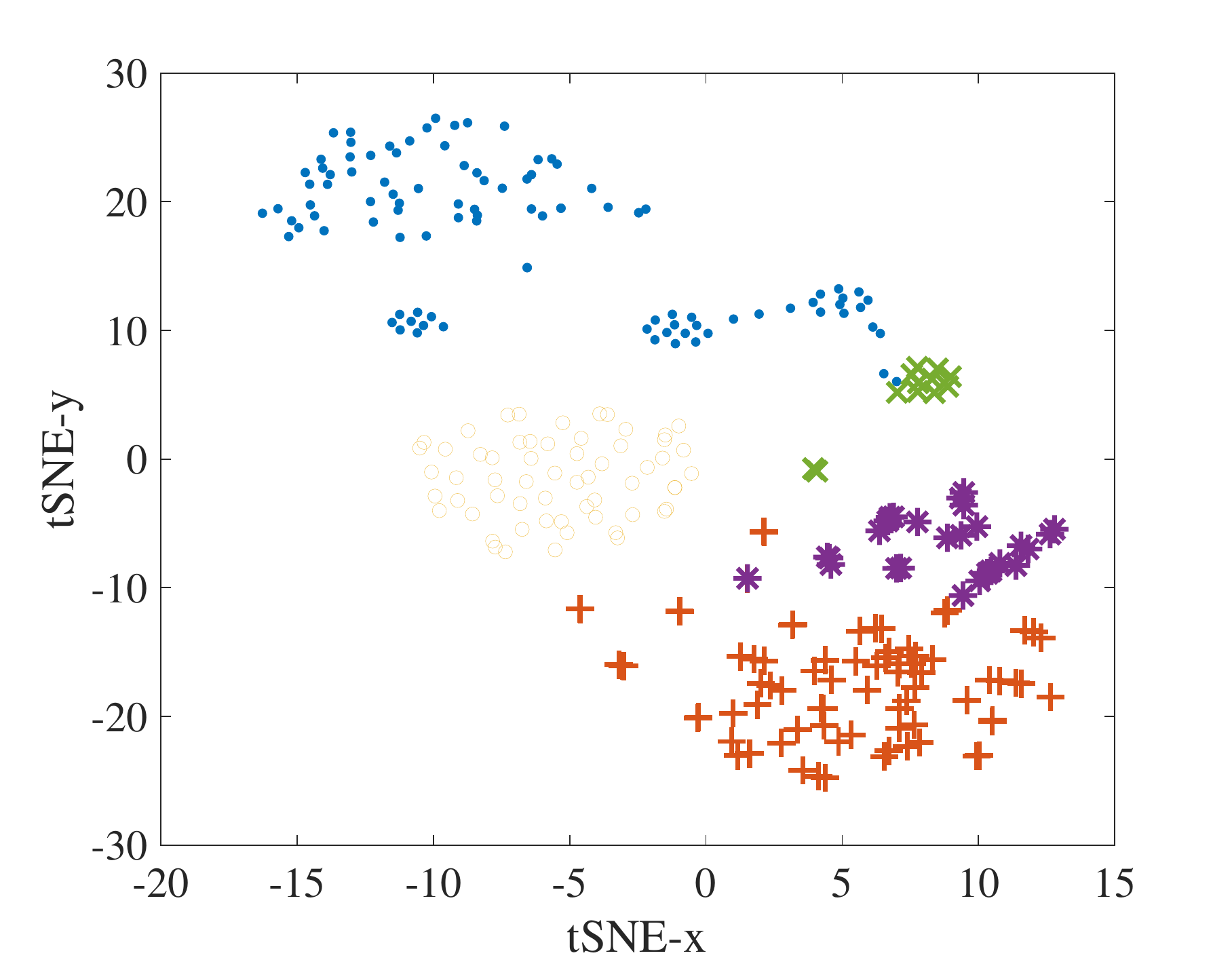}
	\caption{\small Bitcoin clustering: (Left) The $k$-var curve, where the number of token's clusters is found by the Elbow method; (Right) ${\bm t}$-SNE Clustering results.}
	\label{fig:Bittsne}
\end{figure*}
{\section{Dataset and Numerical Results}\label{sec:numeric}}

In this section, numerical results are provided for both Bitcoin data and Ethereum data. The data sets are downloaded from the actual blockchain systems while we also utilize some known pre-processed results based on the real blockchain data.
\subsection{Numerical Results for the Bitcoin Data}
The Bitcoin data we studied comes from the website {\it http://vo.elte.hu/bitcoin}, where the raw Bitcoin data is processed and compressed into several documents; more details could be found in Ref. \cite{timmurphy.org}. In our experiment, we used the document ``txhash.txt'' which is a list of transaction IDs (indexed by the website) and hash pairs to record the hash for each  transactions in chronological order. We intercepted the block data from block number $250,000$ to $252,000$, whose time interval is between ``2013-08-03 12:36:23" and ``2013-08-13 18:11:30". By searching  transaction hash in ``txhash.txt" we found that these transaction IDs range from $21,490,941$ to $22,003,698$. With these IDs, we can search documents ``txin.txt" and ``txout.txt" to find the input addresses and output addresses. Herein, ``txin.txt" records each transaction's input addresses with the amount of Satoshis{\footnote{ The satoshi is currently the smallest unit of the bitcoin currency recorded on the block chain.} and ``txout.txt" records each transaction's output addresses with the amount of Satoshis.  Within this time interval, we can extract in total $512,756$ transactions involving $515,765$ addresses. It is worth noting that at this moment, the addresses may be duplicated. 

Our next step is to pre-process the data by associating the addresses using the heuristic of ``common spend" and ``change address". To process the data by ``common spend", we utilized the data set ``contraction.txt" from the website {\it http://vo.elte.hu/bitcoin}, which is a list of addresses possibly belonging to the same user. The basic idea of this process is that any two input addresses which belong to the same ``user" appear as inputs in the same transaction at least once. After this processing, we have in total $132,431$ transactions and $65,811$ identified unique users (super-addresses) left. To better analyze the details of some key users, we assume that the ``change address" is used rarely and thus we can eliminate the addresses whose occurrence (appears in the transaction input or output) is less than $30$. This process significantly reduces the size of the graph to $3930$ transactions and $279$ users. Notice that some of the users are the combinations of several common spend addresses, and others are the change addresses which appeared in the output of transactions but never appeared in the inputs of a transaction. In Fig. \ref{fig:multiaddr}, we plot the graph representation by using the Gephi software\cite{jacomy2014forceatlas2}, where the continuous graph layout algorithm ForceAltas $2$ is adopted to visualize the graph. Note that herein we do not consider to use any features to define the edge weights. The edge weight is either `0' (no transaction) or `1' (with transaction). From the plot we see that after associating common spend, the graph is well clustered while the dimension of the graph is still large. 
 \begin{figure*}[t!]
	\centering
	\includegraphics[width=17cm]{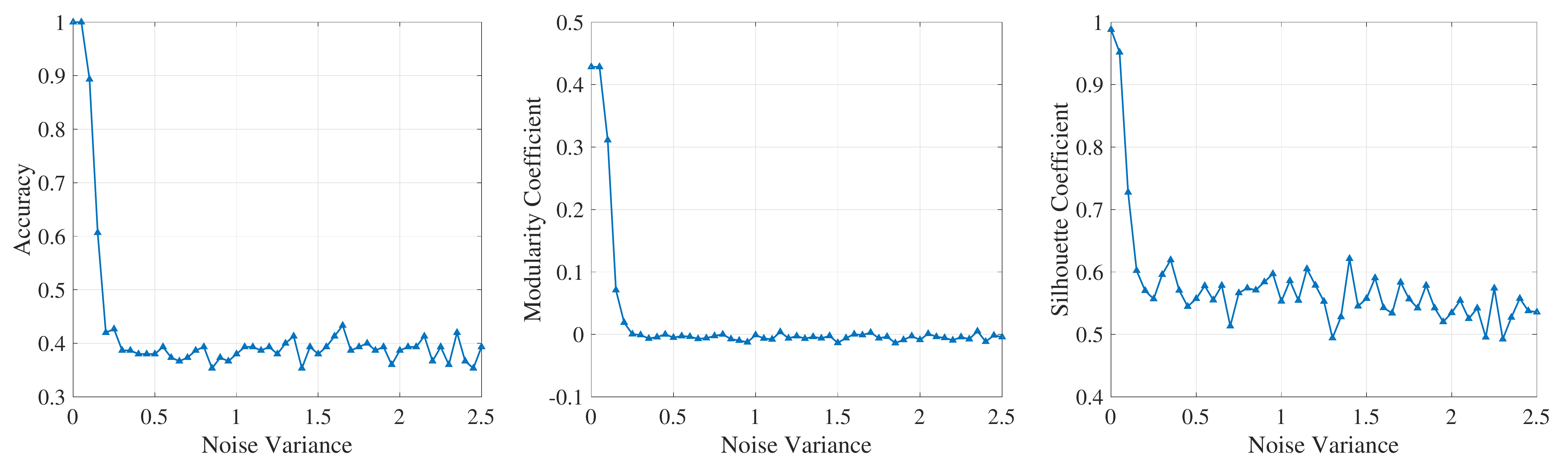}
	\caption{\small How noise affects the recovery of the communities.}
	\label{F:noise}
\end{figure*}

\begin{figure*}[t!]
		\centering
		\includegraphics[width=17cm]{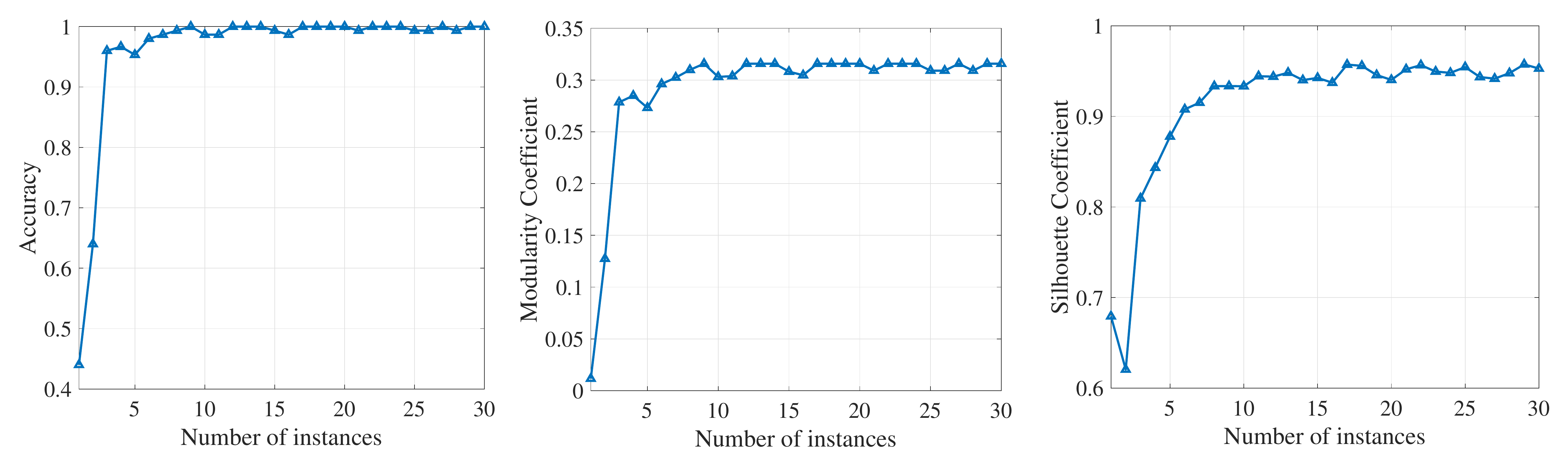}
		\caption{\small Number of instances versus the recovery accuracy.}
		\label{fig:instance}
\end{figure*}

We then evaluate and visualize the community detection results based on the social graph we have defined for the Bitcoin data.
First, we run the Elbow method in Ref. \cite{thorndike1953belongs}  to determine the optimal number of clusters $k$. Fig.~\ref{fig:Bittsne} (left) shows that the ``inflection point" is $k = 5$ and thus we consider there are $5$ clusters in our example. Interestingly, this is also roughly consistent with the results in Fig. \ref{fig:multiaddr} although they have defined different edge weight. We then run Algorithm \ref{alg:1} under a random initialization with $k=5$ and find that the number of nodes in each cluster are $101$, $78$, $60$, $27$, and $13$, respectively. A $2$D visualization of the clustering results are shown in Fig.~\ref{fig:Bittsne} (right), where a machine learning algorithm called ${\bm t}$-SNE\cite{maaten2008visualizing} is used for a nonlinear dimensionality reduction. The results show that the target users are indeed clustered in the compressed $2$D space.
We use two parameters to evaluate the community results. One is the Silhouette score, which combines the two factors of cohesion and resolution to evaluate the clustering results\cite{rousseeuw1987silhouettes}.
The other one is the Modularity score which is usually used to measure the structural of network communities\cite{newman2006modularity}. In this experiment, we have the Silhouette score $0.5871$ and theModularity score $0.3733$. Normally the Silhouette score is at a range of $[-1.0, 1.0]$ and the range of the Modularity score is $[ -0.5, 1.0 ]$. The more two scores approach to $1$, the better quality of the network partition. The Modularity score around $0.3\sim 0.7$ is considered as a good clustering result\cite{newman2004fast}. We track the nodes of the gambling website and find that the gambling website nodes and other $55$ nodes that had transactions with the gambling nodes have been all clustered into the same cluster. This cluster has in total $101$ nodes.

\subsection{Numerical Results for the Ethereum Data}\label{Ethereum_Data}
\subsubsection{Synthetic data test} \label{Synthetic_data}
 To verify our model, we first generate synthetic data to test how the proposed method works for a known graph with given input signal. Specifically, a graph ${\cal G}(N, K, P_a, P_b)$ is generated where $N$ is the number of nodes, $K$ is the number of communities, $P_{a}$ is the probability of node connection within the community and $P_{b}$ is the probability of node connection between communities. We then define the graph filter as $\mathcal{H}(\bm{S}) = (1-\alpha {\bm S})^{L-1}$ where ${\bm S}$ is the Laplace matrix of ${\cal G}$ and $L$ is the order of the graph filter. The graph signal ${\bm y}_t$ is thus generated following \eqref{yt} where we set ${\bm z} = {\bm B} {\bm \alpha}$ with ${\bm B}$ having $R$ non-zero rows and each row having $[{Rd/N}]$ ones where $d$ is the degree of the graph and we generate ${\bm \alpha} \sim {\cal N}(0,  {\bm I})$ to get different instances. Given the structure of the graph and input signal ${\bm z}$, we use three different parameters to evaluate the recovery of the graph. Herein, the Silhouette score and Modularity score have defined before, and the recovery rate is defined as the percentage of nodes that are recovered in the correct cluster.
 
 Fig. \ref{F:noise} shows how noise corrupted in the observed graph signal effects the recovery accuracy. Herein, we set $N=150, P_a=0.89, P_b=0.11, R=15$ and generate $1000$ instances as the input signal. We vary the noise variance to see the recovery performance. The results tell that in the noise-free case, we can $100\%$ recover the communities, while as the noise level increases, the recovery accuracy deteriorates. 
 In Fig. \ref{fig:instance}, we set $N=150, P_a=0.8, P_b=0.2, R=15$ and assume the noiseless case. We vary the number of instances of the input signal to see how it affects the recovery accuracy. Apparently, the results tell that we need to observe sufficient instances to recover the community information. The synthetic data test in Fig. \ref{F:noise} and Fig. \ref{fig:instance} demonstrates that the proposed model can find communities given that the noise in the observation model is not so heavy and we have data for sufficient instances. Motivated by this, in the real data test,  we utilize user-token subscription information in the Ethereum network to find the EoA users' communities.

\begin{figure*}[t!]
		\centering
	    \includegraphics[width=8.2cm]{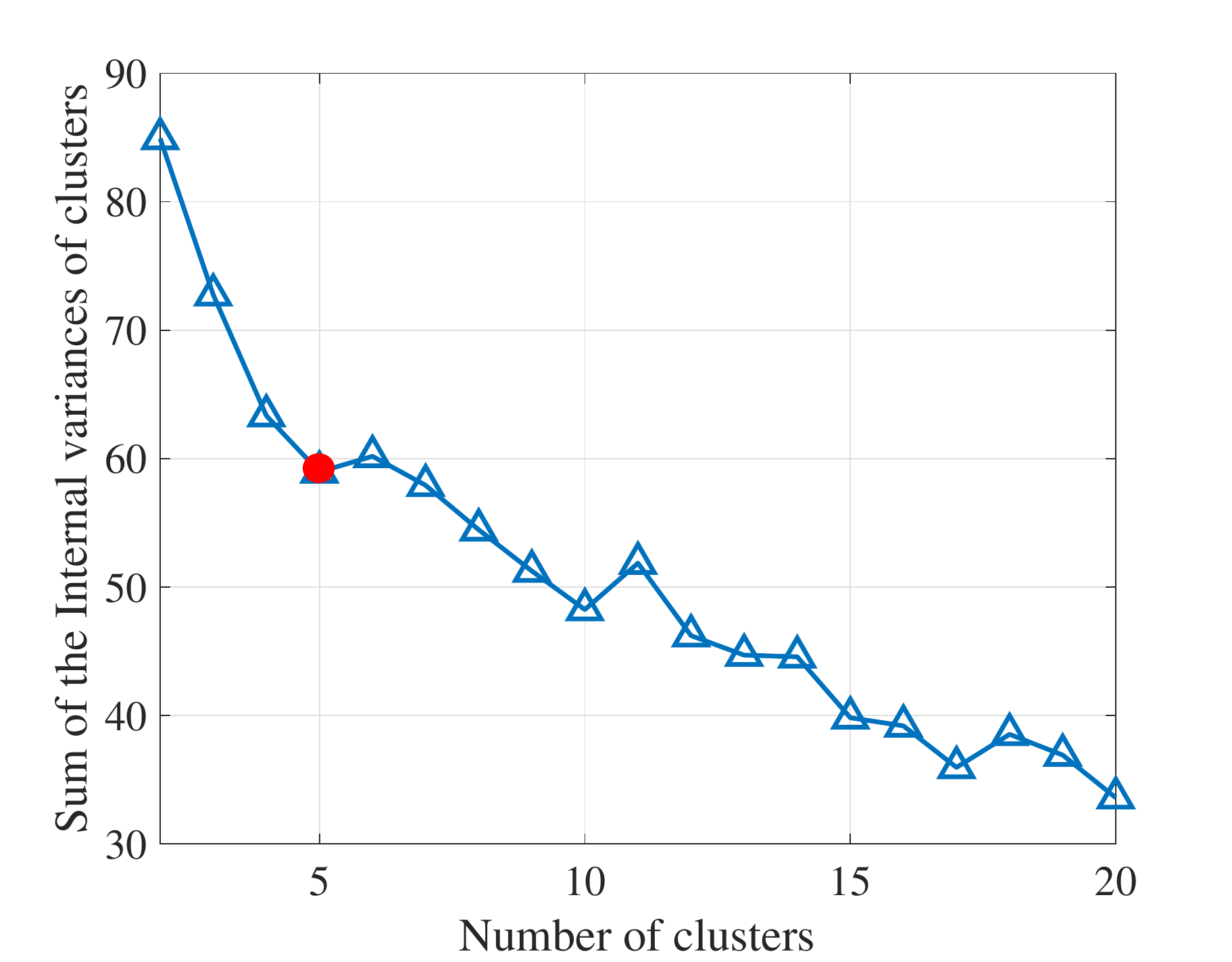}
		\includegraphics[width=8.2cm]{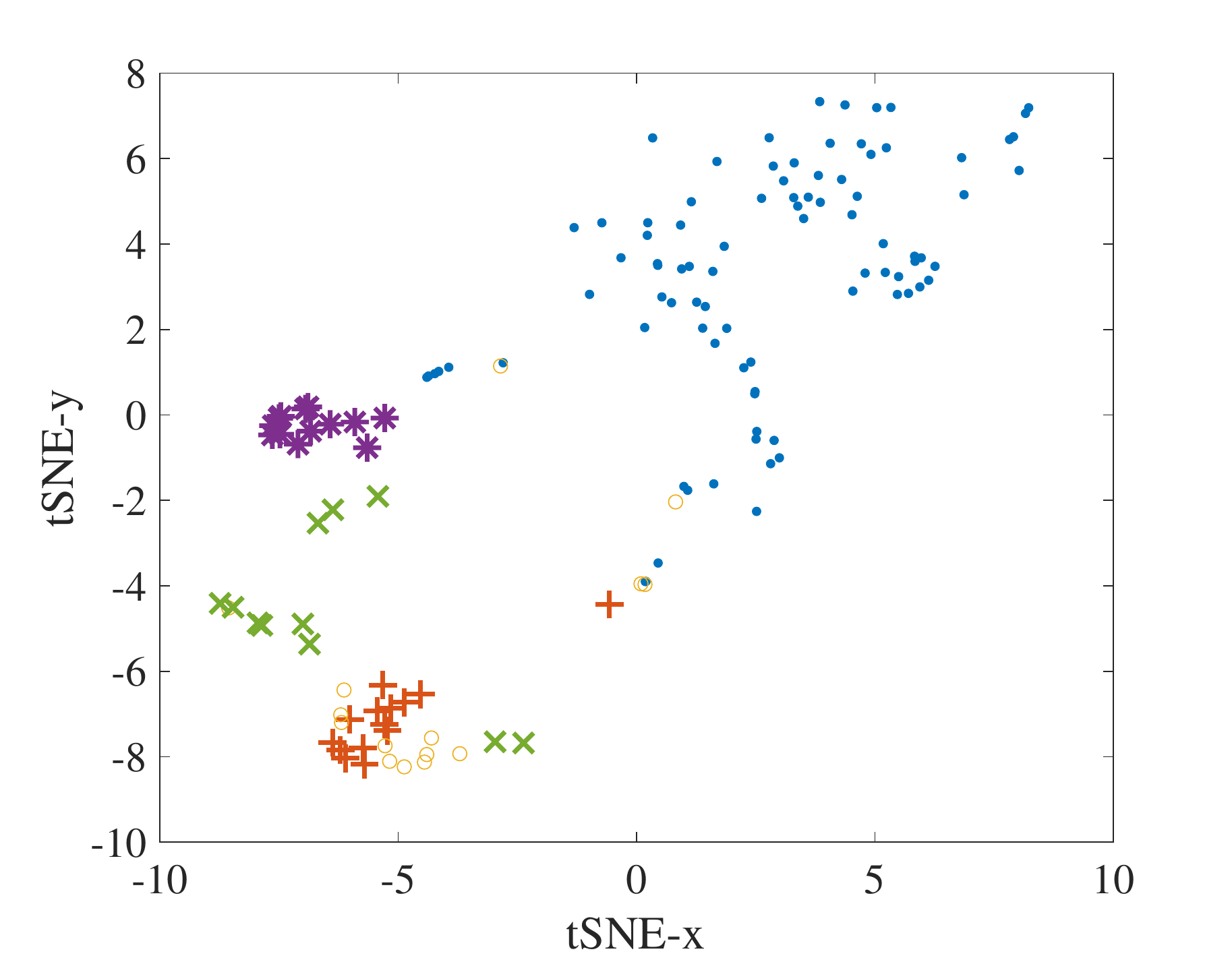}
	    \caption{\small User clustering: (Left) The $k$-var curve, where the number of user's clusters is found by the Elbow method; (Right) ${\bm t}$-SNE Clustering results.}
	    \label{fig:user_cluster}
\end{figure*}
\begin{figure*}[t!]
		\centering
		\includegraphics[width=8.2cm]{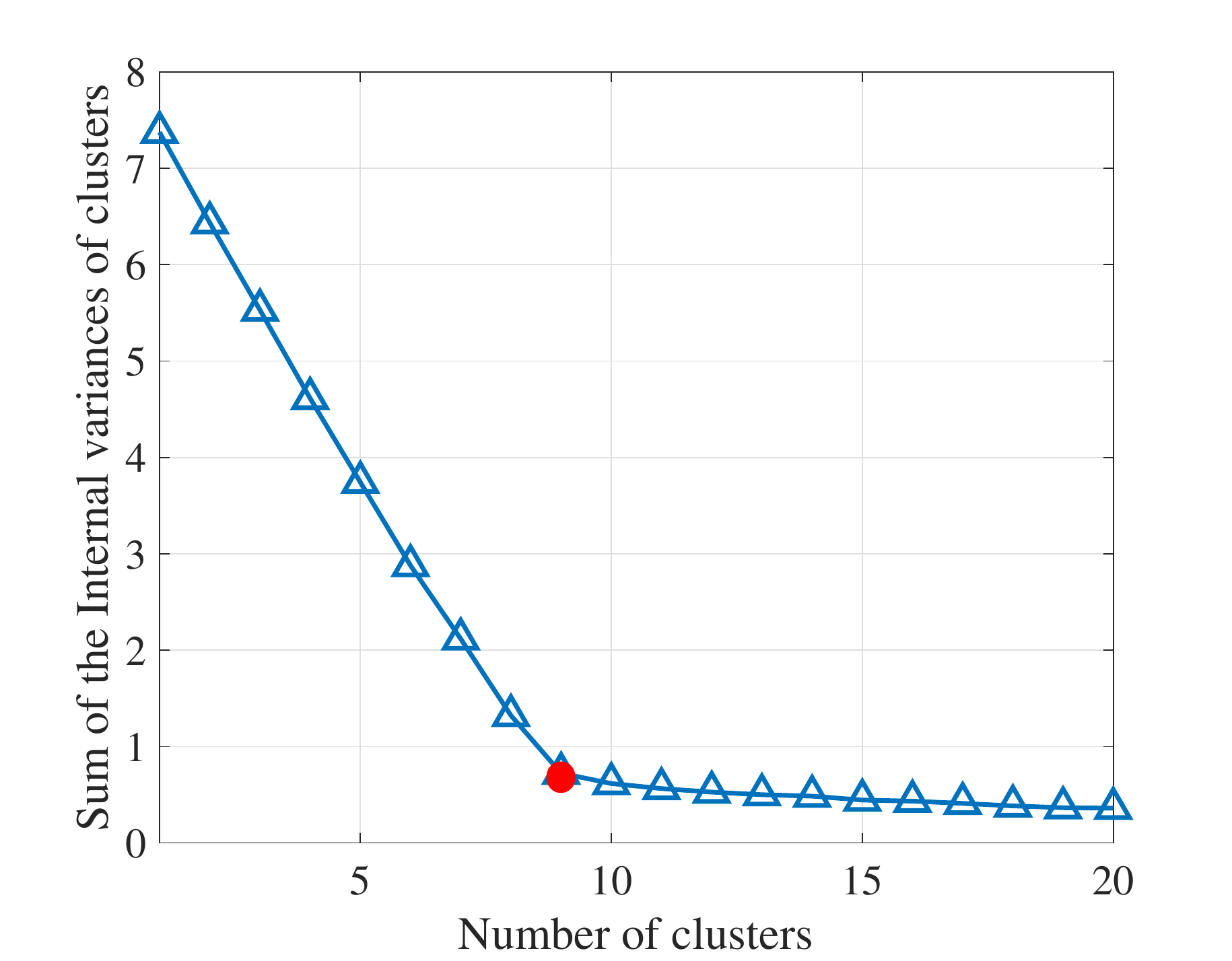}
		\includegraphics[width=8.2cm]{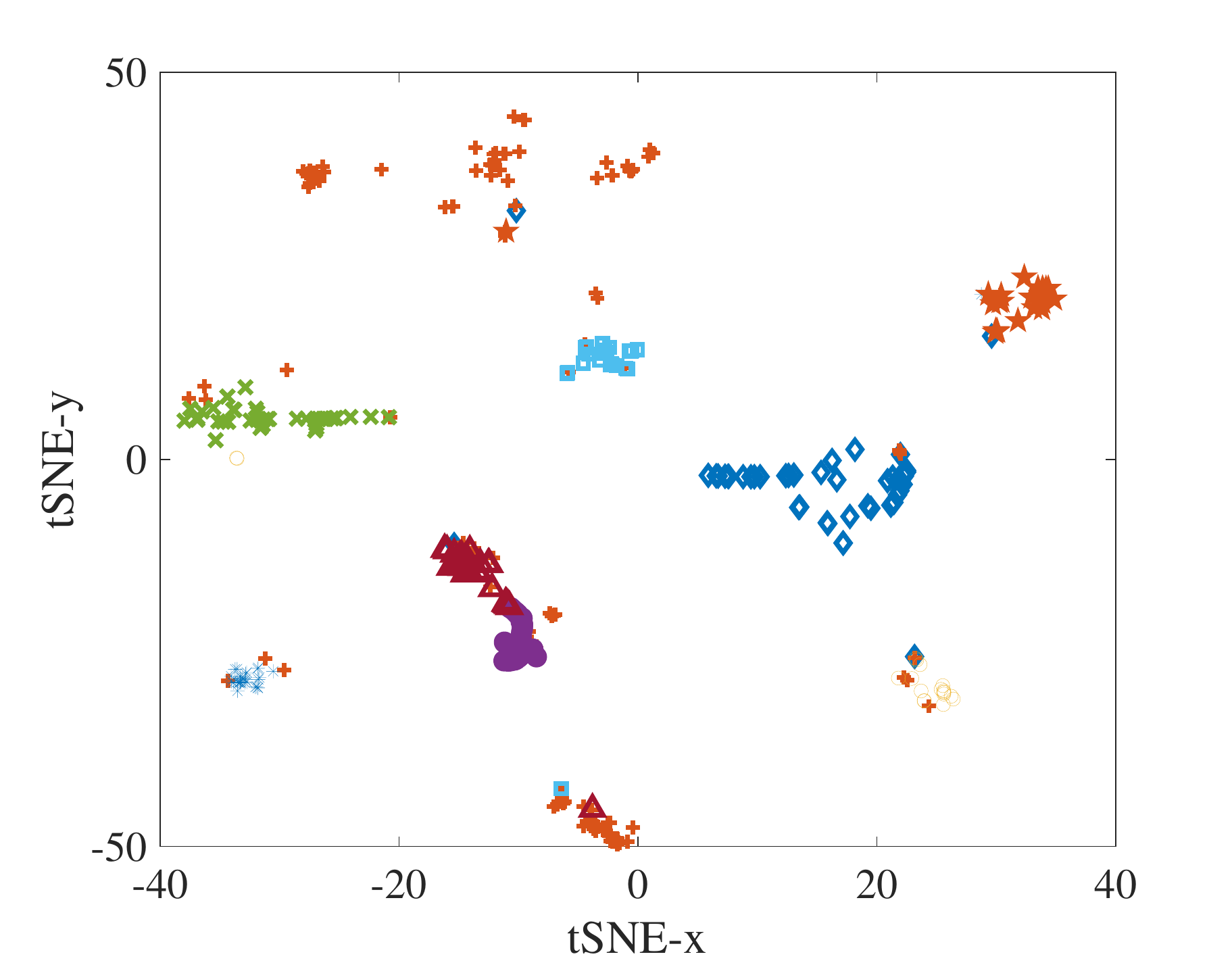}
		\caption{\small Token clustering: (Left) The $k$-var curve, where the number of token's clusters is found by the Elbow method; (Right) ${\bm t}$-SNE Clustering results.}
		\label{fig:token_cluster}
\end{figure*}

\begin{figure*}[htb]
	\normalsize
	\centering
	\includegraphics[width=8.2cm]{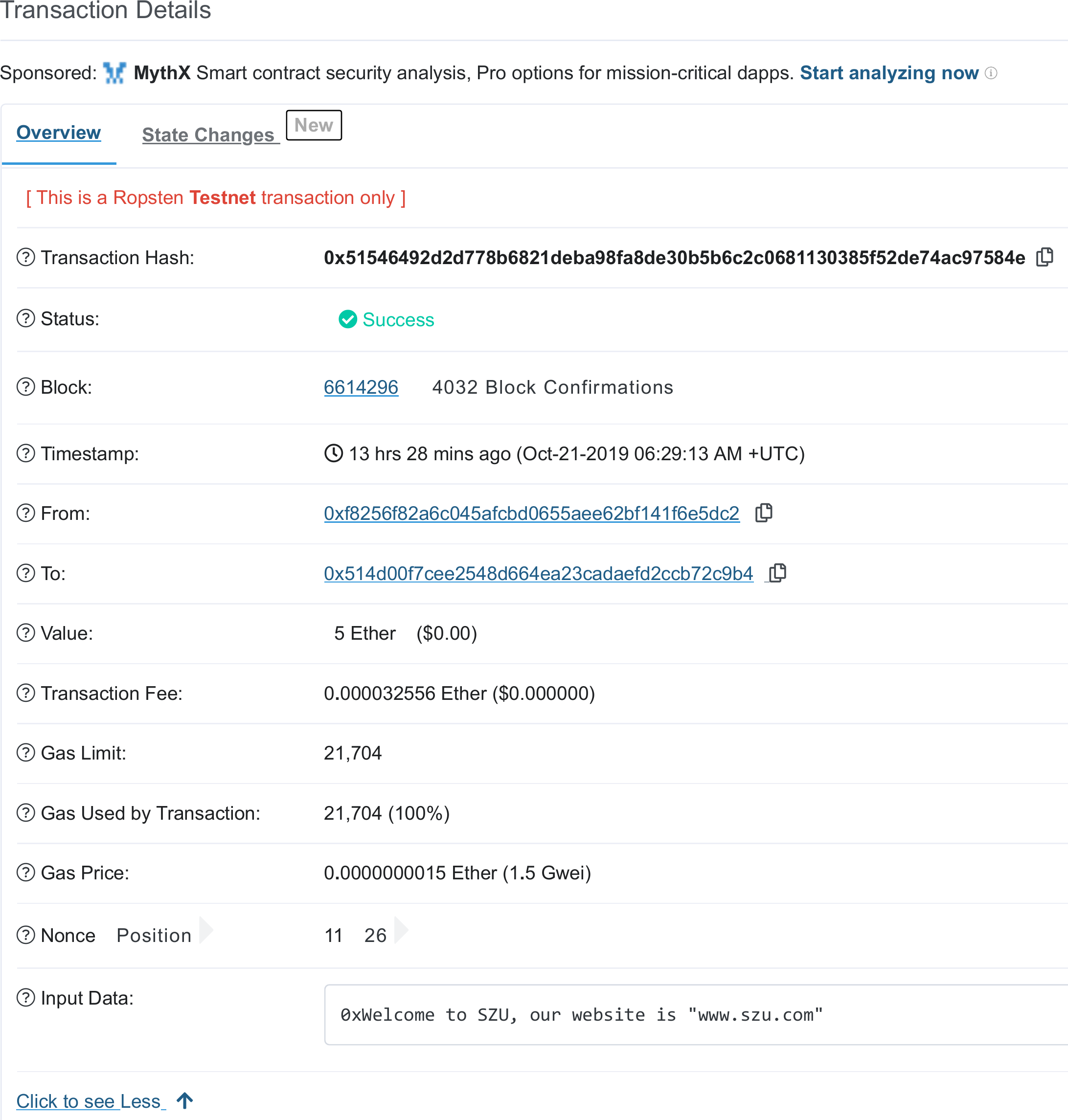}\hspace{6mm}
	\includegraphics[width=8.8cm]{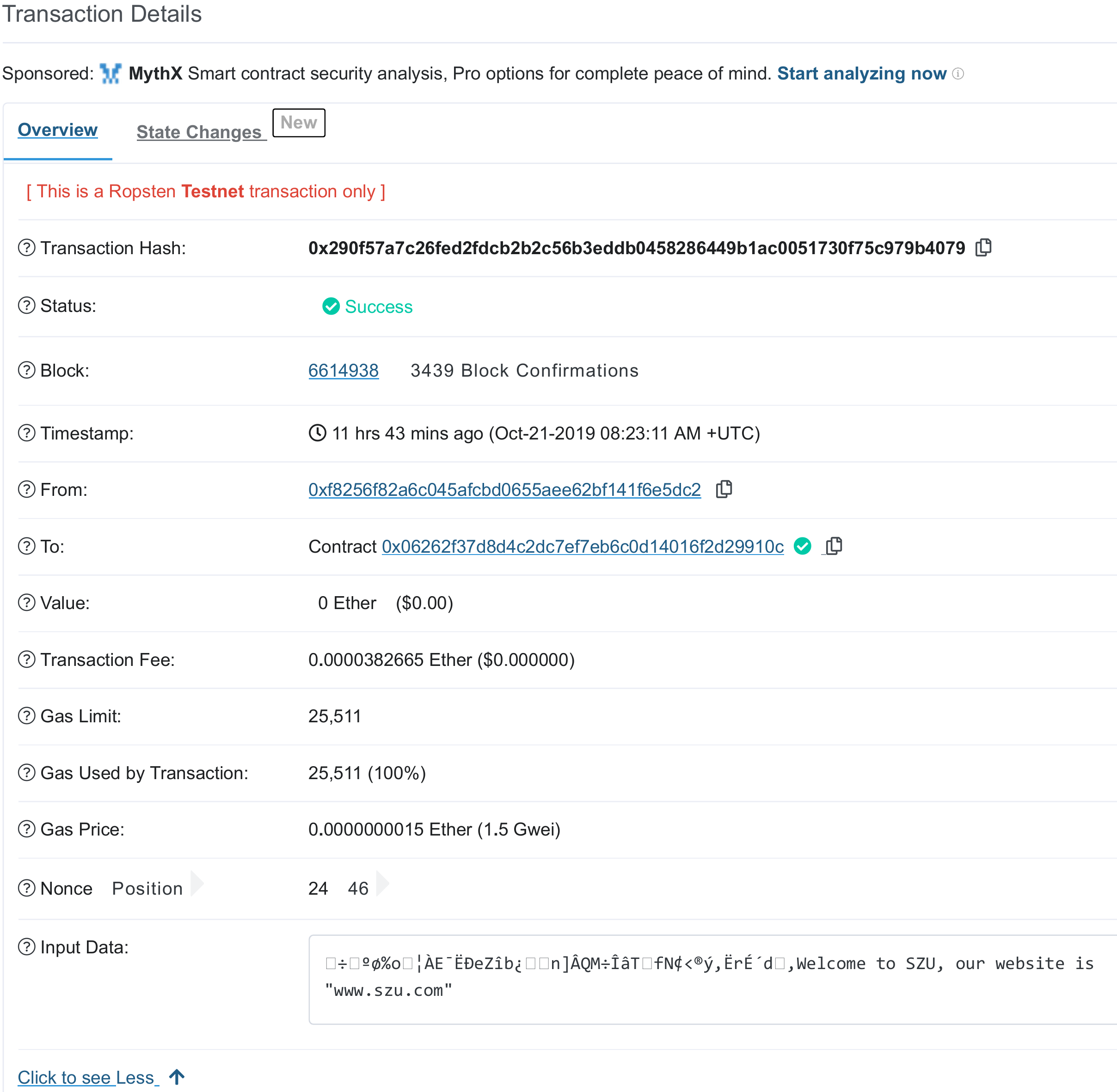}
	\caption{\small View advertisements in Ropsten. These two transactions can be viewed at links: \newline (Left) \href{https://ropsten.etherscan.io/tx/0x51546492d2d778b6821deba98fa8de30b5b6c2c0681130385f52de74ac97584e}{ https://ropsten.etherscan.io/tx/0x51546492d2d778b6821deba98fa8de30b5b6c2c0681130385f52de74ac97584e}, \newline (Right) \href{https://ropsten.etherscan.io/tx/0x290f57a7c26fed2fdcb2b2c56b3eddb0458286449b1ac0051730f75c979b4079}{ https://ropsten.etherscan.io/tx/0x290f57a7c26fed2fdcb2b2c56b3eddb0458286449b1ac0051730f75c979b4079}.}
	\label{F:recomdation}
\end{figure*}

\subsubsection{Real data test} \label{Real_data}
In this part, we focus on the token transactions in the Ethereum network to show a toy example of the data processing. To set up a meaningful example, we screen out the user-token pairs by three steps. In step one, we pick $20$ top market capitalization tokens from website \textit{https://etherscan.io} on the date of July 1st 2019, and record the top $100$ user addresses of each token. 
In step two, we extract aforementioned users' transactions, recorded all the tokens they have owned at the current moment (that would be much more than $20$ tokens mentioned above). After this operation, there are $141$ users with $1837$ tokens in total. Then, in the last step, to maintain an appropriate level of attention for each token, we first delete those users who have subscribed less than $20$ tokens and then delete those tokens which have been subscribed by less than $60$ EoA users. In the end,
we have selected $141$ users who have focused on $21$ tokens to form the bipartite graph. This graph corresponds to a user-token matrix ${\bm A} \in \mathbb{R}^{141 \times 21}$ with entry ${A}_{ij}$ denoting user $i$'s transaction amount on token $j$.  To further process ${\bm A}$, we unify the token value to ETH value by multiplying their currency to ETH on July 1st 2019.
Therefore, in the bipartite graph, the edge weight between the user and the token is defined as the amount of ETH value this user has owned on this token at the current moment. We remark that such a data process can be applied to a much bigger size of data and we can thus analyze a large size graph at one shot. Herein we focus on a small size graph so that we can have a more clear description of the process.

$\bullet$\textit{ {The user clustering:}} After a row normalization and a column normalization to matrix ${\bm A}$, we apply Algorithm~\ref{alg:2} to perform the community detection. In particular, the Elbow method is used to determine the optimal number of clusters for clustering\cite{thorndike1953belongs}. Fig.~\ref{fig:user_cluster} (left) shows that the ``inflection point" is $k = 5$ and thus we consider there are $5$ clusters in our toy example.
Fig.~\ref{fig:user_cluster} (right) shows a $2$D visualization of the clustering results, where ${\bm t}$-SNE is used for a nonlinear dimensionality reduction. We also calculate the Silhouette score and the Modularity score for this case, which are $0.6586$ and $0.5999$, respectively.
	
$\bullet$ \textit{{The token clustering:}} For the previous user clustering process, graph signals $ {\bm y}_{t=1}^T \in \mathbb{R}^{N}$ represents the subscription of all users on token-$t~(t=1,...,T)$, which can be considered as the features of EoA users. To proceed the token clustering, we redefine ${\bm y}_{n=1}^N \in R^{M}$ that can be interpreted as user-$n~(n=1,...,N)$'s subscription on all tokens. We pre-process the data in a similar way as that in the user clustering process and delete the tokens which are not subscribed by any users. The new bipartite graph is thus based on the token-user matrix ${\bm A} \in \mathbb{R}^{1811 \times 141}$. The sample covariance $ \hat{\bm{C}}_{x} $ can also be calculated as in \eqref{eq:Cy}. We then apply Algorithm \ref{alg:2} to cluster the tokens.
Through the Elbow method, Fig. \ref{fig:token_cluster} (left) shows that the ``inflection point" of the token-user graph is $k = 9$. Thus we consider there are $9$ clusters in the token-user graph. By the use of ${\bm t}$-SNE \cite{maaten2008visualizing}, the clustering results are shown in Fig. \ref{fig:token_cluster} (right). The figure displays that the target CA tokens are clustered as $9$ groups. The Silhouette score and Modularity score are $0.5517$ and $0.5027$, respectively. We remark here that for both the user clustering and token clustering ${\bm t}$-SNE results, we could see that most of the nodes are well clustered while there are still few nodes wrongly distributed, owing to the noise in the observation  model.


\subsubsection{Implementation of the On-chain Advertisement}
In this part, we show the implementation of the two on-chain advertisement strategies. Our experiment is done on the Ropsten test net, which is a testing environment for Ethereum. Therein, we have used the MEW module~(c.f., https://www.myetherwallet.com/) to build transactions, the Remix module(c.f., https://remix.ethereum.org/) for deployment contracts and website \textit{https://etherscan.io} to view the block.

In the left of Fig.~\ref{F:recomdation}, we implement how to deliver advertisement in an ETH transaction. We use MEW to build the transaction directly while adding an advertisement message in the input field. Note that to visualize the message we need to convert the string information into hexadecimal. The website \textit{https://etherscan.io} shows us the message in the block. This block will be synchronized to the users' wallet and the wallet will push the message to the target user. Delivering advertisement via smart contract is shown in the right screen of Fig.~\ref{F:recomdation}. Therein, a smart contract transaction is generated by the ICO initiator to send message to a group of users. In this approach, the ICO initiator needs to cooperate with the wallet company to register their token address, as well as pushing the valid advertisement message to target users. Experiments show that both strategies are valid in the testing environment.

{\section{Conclusion} \label{sec:conclusions}}
In this work, we have considered community detection on blockchain networks. We respectively studied Bitcoin and Ethereum networks. In particular, for Bitcoin we defined the social network based on transactions and proposed a modified clustering method for the transaction graph. For Ethereum, a bipartite social graph was defined and a novel low-rank clustering method was adopted to cluster users in this graph. We implemented both methods for real blockchain data, visualized and analyzed the community results. We also demonstrated advertisement strategies for delivering on-chain advertisements in the Ethereum network. Our work verified the possibility of applying community detection in different blockchain networks, given that the observation model is not too noisy and sufficient data is provided. How to reduce the effect of heavy noise would be our next task to conquer.

\vskip 6mm
{\small
\bibliographystyle{IEEEtrans}
\bibliography{reference}
}

\begin{IEEEbiography}
	[{\includegraphics[width=1in,height=1.25in,clip,keepaspectratio]{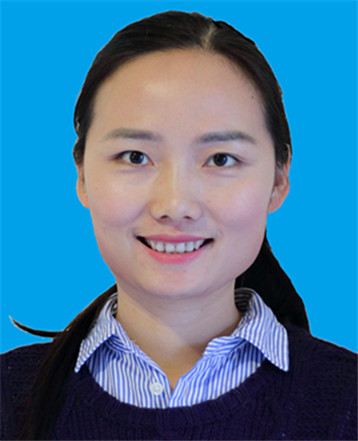}}]
	{Sissi Xiaoxiao Wu} received the B.Eng. degree in electronic information engineering from the Huazhong University of Science and Technology, Wuhan, China, in 2005, the M.Phil. degree from the Department of Electronic and Computer Engineering, Hong Kong University of Science and Technology, Hong Kong, in 2009, and the Ph.D. degree in electronic engineering from the Chinese University of Hong Kong (CUHK), Hong Kong, in 2013.,From December 2013 to November 2015, she was a Postdoctoral Fellow in the Department of Systems Engineering and Engineering Management, CUHK. From December 2015 to March 2017, she was a Postdoctoral Fellow in the Signal, Information, Networks and Energy Laboratory supervised by Prof. A. Scaglione of Arizona State University, Tempe, AZ, USA. Since March 2017, she has been an Assistant Professor at the Department of Communication and Information Engineering, Shenzhen University, Shenzhen, China. Her research interests are in wireless communication theory, optimization theory, stochastic process, and channel coding theory, and with a recent
\end{IEEEbiography}
\vspace{-10mm}
\begin{IEEEbiography}
	[{\includegraphics[width=1in,height=1.25in,clip,keepaspectratio]{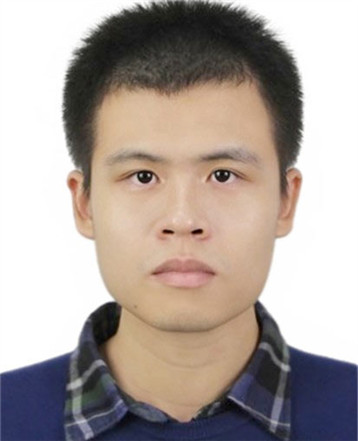}}]
	{Zixian Wu} is currently pursuing a M.Eng degree in communication and information engineering at Shenzhen University. Prior, he received his B.Eng. degree in electronic and information engineering from Shenzhen University, China, in 2019. His research interests are in machine learning, data mining.
\end{IEEEbiography}

\begin{IEEEbiography}
[{\includegraphics[width=1in,height=1.25in,clip,keepaspectratio]{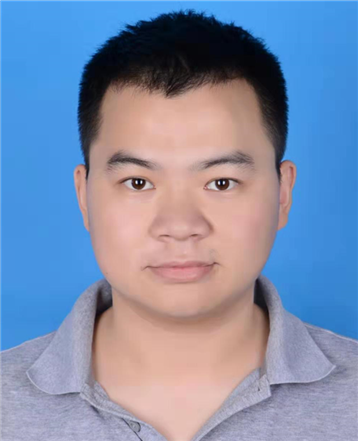}}]
{Shihui Chen} is currently pursuing a M.Eng. degree in Electronics and Communication Engineering at Shenzhen University. Prior, he received his B.Eng. degree in the measurement and control technology and instrument from the Nanchang Institute of Technology, Nanchang, China, in 2017. His research interests are in data mining and blockchain technology.
\end{IEEEbiography}

\begin{IEEEbiography}[{\includegraphics[width=1in,height=1.25in,clip,keepaspectratio]{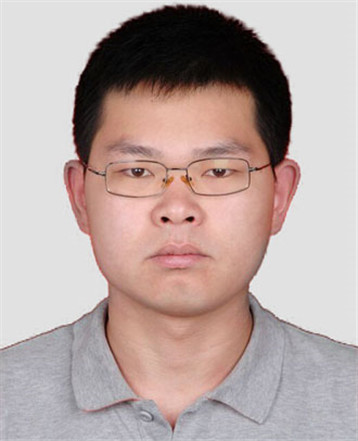}}]
{Gangqiang Li} is currently pursuing a PhD degree in communication and information engineering at Shenzhen University. Prior, he received his B.Eng. degree in electronic engineering from the Henan University of Urban Construction, Pingdingshan, China, in 2014, and the M.Eng. degree in Control Science and Engineering, Shenzhen University, Shenzhen, China, in 2017. His research interests are in machine learning, data mining and distributed protocols.  
\end{IEEEbiography}
\vspace{-10mm}
\begin{IEEEbiography}
[{\includegraphics[width=1in,height=1.25in,clip,keepaspectratio]{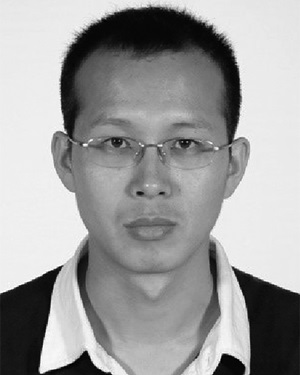}}]
{Shengli Zhang} [corresponding author] received the B.Eng. degree in electronic engineering and the M.Eng. degree in communication and information engineering from the University of Science and Technology of China, Hefei, China, in 2002 and 2005, respectively, and the Ph.D. degree from The Chinese University of Hong Kong, Hong Kong, China, in 2008. After that, he joined the Communication Engineering Department, Shenzhen University. He is currently a Full Professor. From March 2014 to March 2015, he was a Visiting Associate Professor with Stanford University. He is the pioneer of Physical-layer network coding. He has authored or coauthored more than 20 IEEE top journal papers and ACM top conference papers, including IEEE JSAC, IEEE TWC, IEEE TMC, IEEE TCom, and ACM Mobicom. His research interests include physical layer network coding, interference cancellation, and cooperative wireless networks. He served as an Editor for the IEEE TVT, IEEE WCL, and IET Communications. He has also served as TPC member in several IEEE conferences, including IEEE Globecom2016, Globecom2014, ICC2015, ICC2014, WCNC2012, and WCNC2014. 
\end{IEEEbiography}

\end{document}